\documentclass[twocolumn,aps,prl,floatfix]{revtex4}
\usepackage{graphicx,bm}

\begin{document}
\title{Non-collinear spin transfer in Co/Cu/Co multilayers} 
\author{M.D. Stiles}
\affiliation{$^1$National Institute of Standards and Technology,
Gaithersburg, MD 20899-8412}
\author{A. Zangwill}
\affiliation{$^2$School of Physics, Georgia Institute of Technology,
Atlanta, GA 30332-0430}
\date{\today} 

\begin{abstract}
This paper has two parts.  The first part uses a single point of view
to discuss the {\it reflection} and {\it averaging} mechanisms of
spin-transfer between current-carrying electrons and the ferromagnetic
layers of magnetic/non-magnetic heterostructures. The second part
incorporates both effects into a matrix Boltzmann equation and reports
numerical results for current polarization, spin accumulation,
magnetoresistance, and spin-transfer torques for Co/Cu/Co
multilayers. When possible, the results are compared quantitatively
with relevant experiments.

\end{abstract}

\maketitle 

\section{Introduction}
In 1996, Slonczewski \cite{Slonczewski:1996} and Berger
\cite{Berger:1996} pointed out that an electric current that flows
perpendicularly through a magnetic multilayer can exert a torque on
the magnetic moments of the heterostructure. The torque arises because
a polarized electron in a non-magnet feels a large exchange field when
it propagates into a ferromagnet. For at least two distinct reasons,
this interaction induces a transfer of spin angular momentum (and
hence a torque) between the current-carrying electrons and the
ferromagnetic layers of the heterostructure.

One source of spin-transfer, the {\it reflection mechanism}, occurs
because the reflection coefficient for electrons incident on a
magnetic/non-magnetic interface is spin-dependent.  The spin content
of the reflected and transmitted wave functions differ (in general)
so, inevitably, angular momentum is gained or lost to the magnetization in
the immediate vicinity of the interface.  A second source of
spin-transfer, the {\it averaging mechanism}, occurs because the spins
of electrons transmitted into a ferromagnet from a non-magnet precess
around the magnetization of the ferromagnet.  On account of this
precession, the component of the total conduction electron spin
transverse to the magnetization averages to zero when summed over all
electrons.  Since total angular momentum is conserved, the
ferromagnetic moments gain what the electrons lose.

Motivated by theoretical considerations of this sort, and earlier
experimental indications of
current-induced magnetic excitations \cite{others}, groups at Cornell
\cite{Myers:1999,Katine:2000} and Orsay \cite{Grollier:2001} recently
demonstrated that the relative magnetization of the cobalt layers in
Co/Cu/Co trilayer structures (Figure 1) can be switched by
passing an electric current through the structure. The observed
asymmetry of the switching with respect to the direction of current
flow is indicative of the effect of spin transfer torques (rather than an
effect of a current-induced magnetic field).

The theoretical treatment of this problem is complicated by the fact
that the magnetizations of the ferromagnetic layers are necessarily not
collinear \cite{Brataas:2001,Waintal:2000,Hernando:2000}. In this
paper, we use a Boltzmann equation to compute current polarization,
spin accumulation, magnetoresistance, and spin transfer torques in
Co/Cu/Co heterostructures. This approach is restricted to Ohmic
transport, but it permits us to treat situations where the interface
resistance does not necessarily dominate the transport and also where
the layer thicknesses are less than relevant mean-free paths
\cite{waffle}. That 
is, we can treat spacer layers of arbitrary thickness. Our main
results are: (1) spin-flip scattering in the external leads is
sufficient to polarize the current; (2) the two sources of spin
transfer torque identified above combine in a natural way; (3) the
magnitude of the torque depends on the reflection coefficients, the
spin-dependent conductivity of the ferromagnets, and the layer
thicknesses; (4) the dependence of the magnetoresistance on the angle
between the two ferromagnetic magnetization vectors is not exactly
$\cos \theta$; and (5) satisfactory quantitative agreement is found
with the magnetoresistance data of Katine {\it et al.}
\cite{Katine:2000} but not with the data of Grollier {\it et al.}
\cite{Grollier:2001}.

\begin{figure} 
\includegraphics{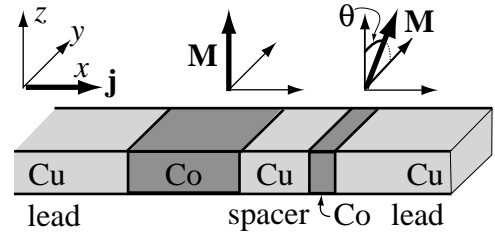} 
\caption{Co/Cu/Co multilayer with non-collinear magnetizations.} 
\label{fig:geom} 
\end{figure}

\section{Observables \& Parameters}

This section defines the observables we use to discuss
transport and spin-transfer. We also 
give the numerical values of the parameters used in our quantitative
calculations for thin Co layers embedded in bulk-like Cu.  
Several of the most relevant observables  
involve incoherent sums of
quantities that are defined quantum mechanically for each electron.
One familiar example is the electron number current density
\begin{eqnarray}
{\bf j}({\bf r}) = -{i \hbar \over 2m}\sum_{\sigma}\left[
\psi^*_{\sigma}\nabla\psi_{\sigma} + {\rm h.c.} \right]
\end{eqnarray}
Less familiar is the current density of spin angular momentum  
\begin{eqnarray}
{\rm Q}({\bf r})=
   - {i\hbar^2 \over 4m}  \sum_{\sigma\sigma'}\left[
        \psi_{\sigma}^* ({\bf r}) 
        \bm\sigma_{\sigma,\sigma'} \otimes
        \nabla \psi_{\sigma'} ({\bf r}) 
      + {\rm h.c.}
    \right] 
\label{eq:Q}
\end{eqnarray}
The gradient of this quantity at any point in space is the local
torque/volume exerted by the electrons on the rest of the system.
Discontinuities are local torques/area exerted by the electrons.
As
the product notation indicates, ${\rm Q}$ is a tensor because 
$\bm\sigma$  has a direction and 
$\nabla \psi$ has a direction. 

It is particularly useful to define a
current polarization vector 
${\bf p}({\bf r})$ by contracting the space part of ${\rm Q}$ with the
number current density:
\begin{eqnarray}
{\bf p}({\bf r})= { 2 \over \hbar}
   {{\rm Q}({\bf r}) \cdot {\bf j}({\bf r})
     \over |{\bf j}({\bf r})|^2 }. 
\label{eq:p}
\end{eqnarray}
For a distribution of electrons, ${\rm Q}({\bf r})$ and ${\bf j}({\bf
r})$ should each be computed separately and then
contracted.  For a completely polarized current, ${\bf p}$ is a unit
vector that points in the direction of the polarization.  
The length of ${\bf p}$ is the up spin
current minus the down spin current, all divided by the total current
for up and down defined with respect to the direction of ${\bf p}$.
Notice that the current polarization and density polarization
(magnetization) need not lie in the same direction or have the same
magnitude.  We discuss such an example below.

We will also have occasion to discuss the voltage drops $\Delta V$
that occur over various portions of the sample. To be precise about
our usage of this symbol, it is important to recall that both the
electric field ${\bf E}$ and gradients of the density deviation from
equilibrium $\delta n({\bf r})$ lead to electric current flow. 
In this context, 
it is usual \cite{Callen} to define an electrochemical potential
\begin{eqnarray}
\overline\mu({\bf r})=  \left[ 2 \pi^2 v_{\rm F} \hbar \over {k_{\rm F}^2 }
\right] \delta n({\bf r}) - eV({\bf r}) 
\end{eqnarray}
as the combination that enters the transport equations.
Here, ${\bf E}= -\bm\nabla V({\bf r})$.  Of course, $V({\bf r})$ and
$\delta n({\bf r})$ are related by the Poisson equation.  But, as far as
the transport equations are concerned, it does
not matter how the two are distributed.  
Therefore, we are free to
choose an approximate solution of Poisson's equation for which there
is no charge accumulation, $\delta n({\bf r})=0$, and interpret $\Delta
\overline{\mu}/e$ as the voltage change $\Delta V$. This is what we
have done in this paper.

On the other hand, the electric field does not couple to the deviation of the
magnetization from its equilibrium value. This is called the spin
accumulation, $\delta {\bf m} ({\bf r})$. Gradients
of the spin accumulation lead to spin currents.  

The numerical results we report in the next section were obtained
by solving a matrix Boltzmann equation (see the Appendix)
appropriate to each portion of the heterostructure shown in Figure 1
(leads, ferromagnets, and spacer
layer). The reflection and averaging mechanisms of spin-transfer
are  
included automatically when we match the solutions together suitably
using a generalization of the boundary conditions described in
Ref. \cite{Penn:1999}. The details will be given elsewhere. 

We make several simplifying approximations which are not intrinsic
to the Boltzmann equation method. We assume that all Fermi surfaces are spheres of the
same size.  Minority and majority electrons in the ferromagnets have
different conductivities due to different Fermi velocities (effective
masses) and different scattering rates.  We also assume that the
interface resistance is due to specular reflection instead of diffuse
scattering.  
We parameterize the reflection amplitudes in terms of
dimensionless parameters $\alpha_\sigma$, chosen to give the correct
interface resistances
\cite{Stiles:2000}, in the form 
\begin{eqnarray} |R_\sigma({\bf k})|^2= {
\alpha_\sigma k_{\rm F}^2 \over \alpha_\sigma k_{\rm F}^2 + k_x^2 },
\label{eq:refl}
\end{eqnarray}
for an electron with wave vector ${\bf k}$ and spin
$\sigma=\uparrow,\downarrow$ incident on an interface with normal
$\hat{\bf x}$.
For simplicity, we choose $R_{\sigma}({\bf k})$ to be real for all the
calculations reported in this paper \cite{real}.
Measured values of the interface resistance for Co/Cu
\cite{Holody:1996,Bass:1999} are consistent with calculated results
from first principles in the specular limit
\cite{Stiles:2000,Schep:1998}, but they are also consistent with
calculations in the diffuse limit \cite{Xia:2001}.

Resistances extracted from
experiments performed at Michigan State \cite{Holody:1996,Bass:1999}
were used to determine the parameters we use to model Co/Cu
structures.  These include the mean
free paths for 
Cu ($\lambda= 110~{\rm nm}$) and 
Co ($\lambda_{\uparrow}= 16.25~{\rm nm}$ 
and $\lambda_{\downarrow}= 6~{\rm nm}$) as well as the
reflectivities for Co/Cu interfaces
($\alpha_\uparrow= 0.051$ and
$\alpha_\downarrow= 0.393$).
The spin-flip mean free path for Cu ($\lambda_{\uparrow
\downarrow}=v_{\rm F} \tau_{\uparrow \downarrow}=2000~{\rm nm}$) was
taken from the 
spin-diffusion length $\sqrt{\lambda \lambda_{\uparrow \downarrow}}$
extracted from a different set of experiments on multilayers grown
electrochemically \cite{Piraux:1998}.  The layer thicknesses were
taken from the experiments done by Katine et al. \cite{Katine:2000}. These are
$t_{\rm Co}(1)= 10.0~{\rm nm} $, 
$t_{\rm Cu}= 6.0~{\rm nm} $, and  
$t_{\rm Co}(2)= 2.5~{\rm nm}$.  

\section{Results}

\subsection{Current polarization by spin-flip scattering}

\begin{figure}
\includegraphics{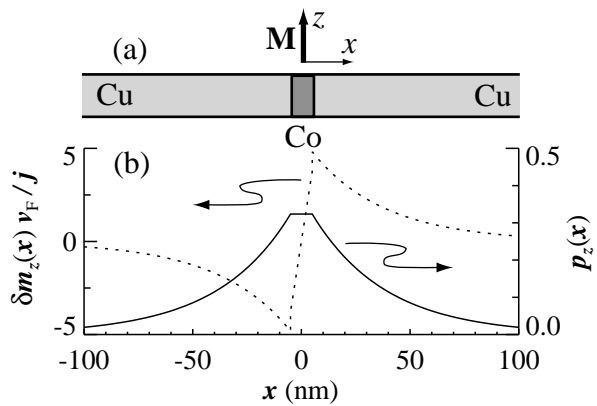}
\caption{Current polarization for a single ferromagnetic layer.
Panel (a):  a thin Co layer embedded between two semi-infinite Cu
leads.  Panel (b): current polarization (solid line)and spin
accumulation (dotted line) for a single Co layer embedded in Cu. The
spin accumulation, defined as a density rather than a magnetization,
is put in a dimensionless, scaled form by dividing by the ratio of the
current to the Fermi velocity.
}
\label{fig:single}
\end{figure}

Inside a ferromagnetic metal like Co, Ohm's law
(${\bf j}_\sigma= \sigma_\sigma {\bf E}$)
guarantees that the current is naturally {\it polarized} ($j_\uparrow \neq j_\downarrow$)
because the conductivities for majority
and minority spin electrons are different ($\sigma_\uparrow \neq \sigma_\downarrow$),
while both spin types feel the same electric field ${\bf E}$.
By the same argument, the current is naturally {\it unpolarized} in a non-magnetic metal
like Cu because $\sigma \uparrow = \sigma_\downarrow = \sigma$.   
However, for a heterostructure like the one shown in panel (a) of
Fig.~\ref{fig:single}---a thin ferromagnetic film sandwiched between
two non-magnetic leads---the steady-state current polarization can deviate
(locally) from its preferred bulk behavior in the presence of
spin accumulation.

To see this, suppose first that spin-flip scattering is absent.
In that case, the number densities
of up and down spin electrons are conserved separately and the two spin
types conduct electricity
in parallel. Moreover, in {\it steady state}, the up and down spin currents (and
the current polarization) are time-independent and spatially uniform everywhere. 
For a layer of Co of thickness $t$ sandwiched
between two Cu leads, each of length $L$, a simple series resistor
model for the two spin channels conducting in parallel gives
the polarization of the current as
\begin{eqnarray}
{ I_\uparrow - I_\downarrow \over I_\uparrow + I_\downarrow } =
{ t (\sigma_\uparrow - \sigma_\downarrow)
  \over L ( 4 \sigma_\uparrow \sigma_\downarrow / \sigma ) +
        t ( \sigma_\uparrow + \sigma_\downarrow ) } .
\end{eqnarray}
This formula shows that the current is unpolarized in the limit that
the leads become infinitely long
($L\rightarrow\infty$). 

Now introduce spin-flip scattering in the leads. The current polarization 
can vary spatially in this case because only the {\it sum} of the up and down
spin currents is conserved. This is shown as the solid curve in panel (b)
of Fig.~(\ref{fig:single}) where ${\bf p}=p_z\hat{\bf z}$.
Note that the
current in the ferromagnet is polarized and the current in the leads (far
from the interfaces) is unpolarized. In between, $p_z(x)$ varies on a
scale set by the spin diffusion length. Therefore, the
presence of spin-flip scattering \cite{sfCo} allows the system to
accommodate as much as possible to the "polarization desires" of
both the ferromagnet and the non-magnet (as determined by their intrinsic conductivities).
Non-zero values of the dashed curve in Fig.~(\ref{fig:single}) identify
portions of space where the spin density deviates from its equilibrium value, {\it i.e.},
spin accumulation. As mentioned earlier, the gradient of this quantity
contributes to the current polarization.

Returning to the solid curve, the fact that $p_z(x)$ is symmetrical around the origin
tells us that the steady state current distribution is equally
polarized on both sides of the ferromagnetic layer. This means that no
torque acts on the magnet.  On the other hand, the non-zero gradient
of $p_z(x)$ elsewhere tells us that distributed torques act throughout
the leads where spin-flip scattering occurs. These torques are equal
and opposite at points which are symmetrically disposed with respect
to the thin film. This means that current flow in this system with a
single ferromagnetic layer induces a bending
stress in the entire structure.  In essence, the conduction electrons
transfer angular momentum from one lead to the other. This interesting
result motivates us to look into the mechanisms of spin transfer in more detail.

\subsection{Spin-transfer by reflection}

The fate of a polarized electron incident on a
ferromagnet depends on the angle between the electron spin moment and the
magnetization direction of the magnet. 
We can encode this effect of quantum mechanical exchange most
concisely using spin-dependent
reflection and transmission coefficients $R_\sigma$ and
$T_\sigma$. This has been discussed qualitatively by Waintal {\it et
al.} \cite{Waintal:2000}. Here, we focus on the scattering state for a
polarized electron in a 
non-magnet ($x<0$) that is incident on a ferromagnet ($x>0$). If the
incident 
electron spin points in an arbitrary direction ($\theta \, \phi$) with
respect to the permanent magnetization, we can write its wavefunction
in the form
\begin{eqnarray}
\psi=
  e^{-i\phi/2} \cos(\theta/2) \left|\psi_{{\bf k}\uparrow}\right\rangle
+ e^{ i\phi/2} \sin(\theta/2) \left|\psi_{{\bf k}\downarrow}\right\rangle. 
\label{eq:psi}
\end{eqnarray}
Here, 
\begin{eqnarray}
\left|\psi_{{\bf k}\uparrow}\right\rangle &=& 
( e^{ikx} + R_{\uparrow} e^{-ikx} ) 
\left|\uparrow\right\rangle
~~~~~x<0\nonumber\\
&=&~~~~~~~~~~~ T_{\uparrow} e^{ik_\uparrow x}
\left|\uparrow\right\rangle
~~~~~x>0\nonumber\\
\left|\psi_{{\bf k}\downarrow}\right\rangle &=& 
( e^{ikx} + R_{\downarrow} e^{-ikx} )
\left|\downarrow\right\rangle
~~~~~x<0\nonumber\\
&=&~~~~~~~~~~~ T_{\downarrow} e^{ik_\downarrow x}
\left|\downarrow\right\rangle
~~~~~x>0
\label{eq:eigen}
\end{eqnarray}
are scattering states in a majority/minority basis.  Inserting
Eq.~(\ref{eq:psi}) into Eq.~(\ref{eq:p}) gives the incident current
polarization as
\begin{eqnarray}
{\bf p}_{\rm inc} = 
(\sin\theta\cos\phi,\sin\theta\sin\phi,\cos\theta) .
\end{eqnarray}

It is straightforward (but tedious) to compute the corresponding
quantities ${\bf p}_{\rm refl}$ and ${\bf p}_{\rm tr}$ from the
transmitted and reflected waves generated by Eq.~(\ref{eq:psi}). We omit
them here and focus instead on the extreme case where $R_{\uparrow}=1$
and $R_{\downarrow}=0$ for an incident electron with a spin pointed in
the $y$ direction (the magnetization in the $z$ direction).  In
this case, the incident spin current polarization is ${\bf p}_{\rm
inc}=(0,1,0)$.  Only up spins are reflected, 
so the reflected spin current polarization is ${\bf p}_{\rm refl}=(0,0,1)$. 
Only down spins are transmitted, so the
transmitted spin current polarization is ${\bf p}_{\rm
tr}=(0,0,-1)$.  

Note first that the $z$ component of  ${\bf p}_{\rm inc}$ is the same
as the $z$ component of ${\bf p}_{\rm refl}+{\bf p}_{\rm tr}$. The numerical
value happens to be zero in this case, but the stated equality is a
general result. Nothing very interesting happens to the component of the
electron spin that is parallel to the quantization axis of the ferromagnet.
By contrast, the transverse component of
the spin angular momentum does change. From Newton's law (and Ehrenfest's
theorem), this is possible only
if the magnetization exerts a torque on the
conduction electron spins. For other
angles and other reflection amplitudes, the amount of transferred
angular momentum is more complicated, but it is non-zero in general.

From the sentence below
Eq.~(\ref{eq:Q}) and using Eq.~(\ref{eq:p}), the
torque exerted on the permanent magnetization at $x=0$ due to this reflection
mechanism is
\begin{eqnarray}
{\bf N}_{\rm R}= A{\hbar\over 2}
\left( {\bf p}_{\rm inc} j_{\rm inc}  - {\bf p}_{\rm tr} j_{\rm tr} 
     - {\bf p}_{\rm refl} j_{\rm refl} \right)_{\perp}.
\label{eq:ts}
\end{eqnarray}
Here, $A$ is the cross sectional area of the interface and the currents $j_{\rm
inc}$, $j_{\rm refl}$ and $j_{\rm tr}$ are taken to be positive. The
subscript $\perp$ reminds us that this vector is transverse to the
magnetization.  We have chosen ${\bf M}=M\hat{\bf z}$, so the
torque lies in the $x-y$ plane---specifically, the $y$-direction for our simplified
example.

\subsection{Spin-transfer by averaging}

The averaging mechanism of spin transfer is also a consequence of the
exchange interaction. But, it is completely distinct from the reflection mechanism.
To see this, observe first that the incident electron wavefunction Eq.~(\ref{eq:psi})
in the non-magnet is a coherent superposition of two degenerate spinors with the same
wave vector. When this electron enters the ferromagnet, the majority
and minority spin components at the Fermi surface no longer share the
same wave vector. As a result, the electron spin precesses rapidly in
real space \cite{precession}. The spatial precession frequencies vary
rapidly over the Fermi surface so, when we sum over all
current-carrying electrons, the transverse component of the total
conduction electron spin averages to zero.  In other words, 
an ensemble of electrons that enters a ferromagnetic layer with a {\em
non-zero} transverse component of the current polarization, exits the
layer with {\em zero} transverse component.
From the change, the torque the ensemble
exerts on the permanent magnetization is ${\bf N}_{\rm
A}=\frac{1}{2}\hbar A j_{\rm tr}({\bf p}_{\rm tr})_{\perp}$.
This ``averaging" torque cancels part of the ``reflection" torque Eq.~(\ref{eq:ts}) so
the net spin-transfer
torque is
\begin{eqnarray}
{\bf N}= {\bf N}_{\rm R}+{\bf N}_{\rm A}=A{\hbar\over 2}
\left( {\bf p}_{\rm inc} j_{\rm inc}  
- {\bf p}_{\rm refl} j_{\rm refl} \right)_{\perp}.
\label{eq:torque}
\end{eqnarray}

The net torque manifests itself in a discontinuity in the transverse angular-momentum
current; the latter is zero inside the ferromagnet.
For an electron ``beam'' with current density $j$, this 
torque is 
\begin{eqnarray} 
{\bf N}=Aj{\hbar\over 2 } 
[1-R_\uparrow R_\downarrow] 
(\sin\theta\cos\phi,\sin\theta\sin\phi,0)
\label{eq:torque_electron} 
\end{eqnarray} 
if each electron is described by Eq.~(\ref{eq:psi}). 
We remind the reader that $R_\uparrow$ and $R_\downarrow$ are both real in
our calculation. Also, since both reflection and averaging 
contribute to the torque, there can be extreme cases where only one or the other
contributes. For example, only the
reflection mechanism contributes if $R_{\uparrow}$=1 and $R_{\downarrow}=0$.
Conversely, only the averaging
mechanism contributes if $R_{\uparrow}=0$
and $R_{\downarrow}=0$. Both 
happen to give the same numerical result for these particular
cases ($R_{\uparrow}R_{\downarrow}=0$ in
Eq.~(\ref{eq:torque_electron}) for both cases). Finally, 
it is worth noting that the product 
$R_{\uparrow}R_{\downarrow}$ in Eq.~(\ref{eq:torque_electron}) is closely
related to the {\it mixing conductance} $G_{\uparrow \downarrow}$ used
in the ``circuit'' theory of Ref.~\cite{Brataas:2001}.

\subsection{Non-collinear transport in a trilayer}

\begin{figure}
\includegraphics{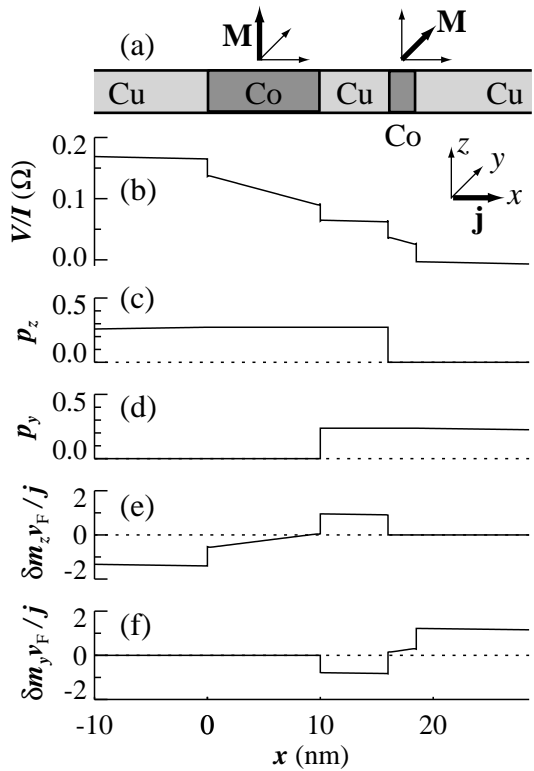}
\caption{Voltage, current polarization, and spin accumulation for a trilayer.
Panel (a): a
heterostructure with two
Co layers, an interposed Cu layer, and two 
semi-infinite Cu leads.  Electron current flows
in the $x$-direction and the left magnetization is in the $z$ direction
and the right is in the $y$ direction.
Panel (b): the voltage drop
(electrochemical potential) through the structure.  Panels (c) and (d):
$z$ and $y$ components of the current polarization, respectively. Panels
(e) and (f): $z$ and $y$ components of the spin accumulation (see
Fig~\protect\ref{fig:single}), respectively.}  
\label{fig:dist}
\end{figure}

We now apply all the above to a trilayer structure modeled after the
experiments of \cite{Katine:2000} and \cite{Grollier:2001}. The
one-dimensional geometry is shown schematically in panel (a) of
Fig.~\ref{fig:dist}.  For simplicity, we first consider a situation where
the magnetization of the left ferromagnet points along $\hat{\bf z}$
and the magnetization of the right ferromagnet points along $\hat{\bf
y}$.  We omit spin-flip scattering in the spacer layer because its
thickness is small compared to $\lambda_{\uparrow \downarrow}$.
Solving the Boltzmann equation, we see from panel (b) that the
``voltage drop'' is largest across the interfaces (because the
interface resistance is large) but not at all negligible across the
layers themselves.  The relative slopes of the lines in the Co and Cu
layers reflects their relative resistivities.

Panels (c) and (d) of Fig.~\ref{fig:dist} show the current
polarization along the magnetizations directions of the left and right
ferromagnets respectively.  Both are discontinuous at the interface
with the misaligned ferromagnet.  This discontinuity is the origin of
the torque exerted on the respective magnetizations. Moreover, as in
the single-layer problem, ${\bf p}(x)$ decreases (too slowly to be
seen in this plot due to the long spin-diffusion length) toward
zero in each lead as $x \rightarrow \pm \infty$. This again
corresponds to a distributed torque in each lead.  Thus, for electron
flow from left to right through the multilayer, the
conduction electrons extract angular momentum from the
lattice of the left lead (by spin-flip scattering) and deposit
an equal amount of angular momentum into the magnetization of the right
ferromagnet. A similar transfer
occurs between the right lead and the magnetization of the left
ferromagnet. The transfers 
are in the same direction, as pointed out in
Ref. \cite{Slonczewski:1996}, so that the current induces the two Co
layers to ``pinwheel'' in the same direction.

Panels (e) and (f) show the spin accumulation along the respective
magnetization directions that is required for consistency with the
calculated current polarizations.  The obvious discontinuities in spin
accumulation across the interfaces are due to the large spin
dependence of the interface resistances.  From panels (c) and (d), the
polarization of the current in the Cu spacer layer is roughly in the
$\hat{\bf z}+\hat{\bf y}$ direction inside the spacer layer, while
from panels (e) and (f), the polarization of the density in the Cu
spacer layer is roughly in the $\hat{\bf z}-\hat{\bf y}$ direction.
The two are not collinear.

\subsection{Angular dependence of resistance and torque}

\begin{figure}
\includegraphics{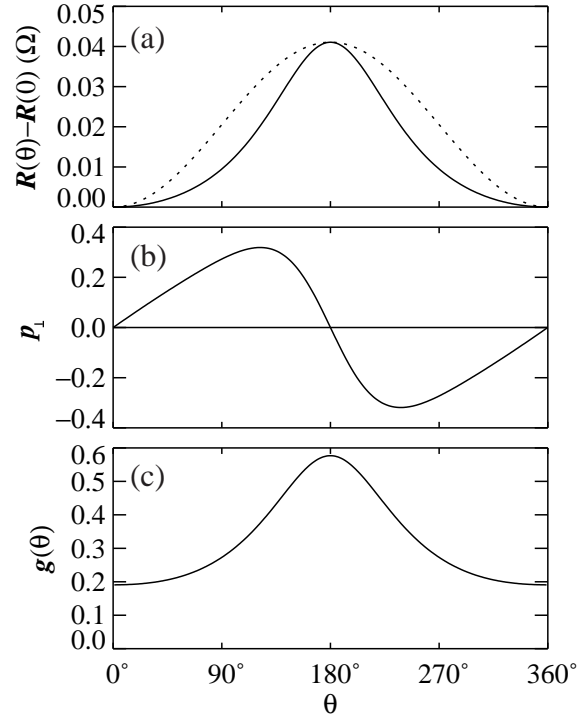}
\caption{Magnetoresistance and torque.  Panel (a) shows the change in
resistance as a function of the relative angle between the two
magnetizations (solid curve).  The dotted curve is proportional to
$1-\cos\theta$.  The resistance has been computed from the
values in the text and a cross-sectional area of
$\pi~65^2$ nm$^2$.  Panel (b) shows the transverse current
polarization on the right ferromagnetic layer.  Panel (c) shows the
same quantity divided by $-\sin\theta$.}
\label{fig:ang}
\end{figure}

For the structure illustrated in Fig.~\ref{fig:dist},
Fig.~\ref{fig:ang} shows the dependence of the resistance and the
torque on the angle $\theta$ between the magnetizations of the two
ferromagnets.  Panel (a) shows that while the magnetoresistance varies
roughly like $1-\cos\theta$, there are significant deviations.
Several authors \cite{qm_ang_dep} find similar deviations using a
fully quantum mechanical treatment. Our results show that
non-sinusoidal behavior occurs already at the semi-classical level if
spin non-collinearity is treated properly.

Our computed value of $R(180)-R(0)$ is about half of the value
measured by Katine et al. \cite{Katine:2000}. Possible sources of this
discrepancy are (1) experimental uncertainty in the multilayer
cross-sectional area (on the order of 40\%); (2) material differences
in the structures grown at Cornell and Michigan State; and (3) the
treatment of the leads in the calculation.  In our results, the leads
have a higher resistance for parallel alignment than for antiparallel
alignment.  The current is largely unpolarized in the latter case, but
not in the former, and there is extra resistance associated with the
spin-flip scattering that polarizes the current.  We suspect that the
wider leads used in the experiment would reduce this effect leading to
better agreement between the calculation and the measured results.
Even though a series resistor model is not justified for layers
thinner than the relevant mean free paths, we find from such a model,
with no resistance in the leads, a factor of two increase in the
difference in resistance, in much better agreement with the measured
result.

We have carried out similar calculations to compare with the results
of Grollier et al. \cite{Grollier:2001}.  That comparison is much less
satisfactory.  Using their experimental geometry and the same
transport parameters, we compute $R(180)-R(0)$ to be about
0.006~$\Omega$. This is much larger than the experimental value of
about 0.001~$\Omega$. Moderate changes in the Co and Cu layer
thicknesses do not change the results very much because the interfaces
dominate the physics. We could bring the calculation into agreement if
there was much less spin-dependence to the interface resistance or if
the cross-sectional area of the multilayer was much different than the
quoted value.  In addition to the possible sources of error discussed
above, it is also possible that the ferromagnetic layers are not
uniformly magnetized in either the parallel or the antiparallel state.

Panel (b) of Fig.~\ref{fig:ang} shows the transverse part of the
current polarization at the interface with the left ferromagnetic
layer.  Again, 
this curve deviates significantly from simple $\sin\theta$ behavior.
The deviation is highlighted in panel (c) which shows
\begin{eqnarray}
g(\theta)= -p_\perp(\theta)/\sin\theta .
\end{eqnarray}
We find that the deviations for the transverse current polarization
track the deviations for the magnetoresistance as we vary material
parameters.  These deviations are quite pronounced, even for
completely symmetric structures.  The maxima in the transverse
currents do not occur for perpendicular alignment of the
magnetizations, but rather happen nearer to antiparallel alignment.  For a
symmetric structure with magnetizations perpendicular to each other,
the current polarization is only $45^\circ$ away from the
magnetization.  The current polarization becomes perpendicular to the
magnetizations as they become antiparallel, but the amount of
polarization decreases to zero in that limit.  This is significant
because the torque on the left ferromagnetic layer is proportional to
the transverse part of the spin current incident on the interface.

The magnitude of the torques we compute are consistent with those that
cause reversal in experiment, but direct comparison is difficult. As
has been pointed out by others \cite{Katine:2000,reversal}, it is not
simply a matter of computing when some energy barrier is overcome.
The torque is zero in both the parallel and antiparallel
configurations so fluctuations away from these orientations can be
amplified by the current-induced torque.  The other sources of
torque---magnetostatics, magnetocrystalline anisotropy, and external
fields---lead to precession.  The damping tends to reduce the
amplitude of the precession, counteracting the effects of the
current-induced torque.  At some point, the current becomes high
enough that a complicated reversal occurs.

\section{Summary}

We have used a matrix version of the Boltzmann equation to study
perpendicular transport in submicron multilayers where the
magnetizations of the ferromagnetic layers point in different
directions.  Spin-flip scattering in the leads ensures that
a polarized current flows.  The boundary conditions for the Boltzmann
equation incorporate the reflection and averaging mechanisms of spin
transfer discussed by Slonczewski and
Berger. As a result, the conduction electrons and the magnetic moments of the
ferromagnets exert torques on one another. Using material parameters
extracted from experiment, we computed the magnetoresistance, spin
accumulation, current polarization, and magnetization torques for
Co/Cu/Co structures similar to those used in experiments. The
transport data were compared quantitatively with data obtained at
Cornell and Orsay and reasons were suggested to explain some
discrepancies between theory and experiment. The magnitudes of the
computed torques were comparable to the torques that induce
magnetization reversal in the experiments.

\section{Acknowledgments}
A.Z. gratefully acknowledges support from National Science Foundation
grant No. DMR-9531115.   

\appendix

\section*{The Boltzmann equation}

The semi-classical Boltzmann equation is a standard approach to
transport calculations that lies between a fully quantum calculation
and classical drift-diffusion theory.  In the portions of space
occupied by ferromagnetic layers, we use Fert's ``two-current''
description \cite{Valet:1993} where $f_\uparrow({\bf k},{\bf r})$ and
$f_\downarrow({\bf k},{\bf r})$ describe the occupancy of up and down
spin electrons in the phase space volume $d{\bf r}\,d{\bf k}$.
Specifically, we restrict the wave vectors ${\bf k}$ to lie on the
Fermi surface and let $g_{\sigma}({\bf k},{\bf r})$ denote the {\it
change} in the occupancy of electrons of spin type $\sigma$ that
occurs when we apply an electric field ${\bf E}$ to the system. In the
linearized relaxation-time approximation, $g_{\sigma}({\bf k},{\bf
r})$ satisfies the stationary Boltzmann equation
\begin{eqnarray}
{\bf v}_{{\bf k}\sigma} \cdot
{ \partial g_\sigma({\bf k},{\bf r}) 
\over \partial {\bf r} }
- e {\bf E} \cdot
{\bf v}_{{\bf k}\sigma} 
&=&
- { g_\sigma({\bf k},{\bf r}) - 
    \overline{g}_\sigma({\bf r}) \over \tau_\sigma }
\label{eq:be_fm}
\end{eqnarray}
where ${\bf v}_{\bf k}$ is the velocity of a state on the Fermi
surface, $\tau_\sigma$ is the spin-dependent relaxation time, and
$\overline{g}_\sigma({\bf r})$ is the average of $g_\sigma({\bf
k},{\bf r})$ over the Fermi surface.

We assume that non-magnetic leads carry current into or out of each
ferromagnet.  In that case, each lead can share the quantization axis
defined by the ferromagnet to which it is connected. 
A spin-flip
scattering term
\begin{eqnarray}
{g_\uparrow({\bf k},{\bf r})-g_\downarrow({\bf k},{\bf r})
\over \tau_{\uparrow \downarrow}}
\label{eq:sf}
\end{eqnarray}
for both spin types is included in Eq.~(\ref{eq:be_fm}) where
appropriate.

The non-magnetic spacer layer must be treated differently because the
non-collinearity of the two ferromagnets induces a spin polarization
in the spacer whose direction generally varies in both real space and
reciprocal space.  To treat this situation, we use a $2 \times 2$
Hermitian occupation matrix ${\bf f}({\bf k}, {\bf r})$ in place of
the functions $f_{\sigma}({\bf k},{\bf r})$ \cite{BaymPethick:1991}.
In particular, at a point where the natural spin quantization axis
points in the direction $(\theta \, \phi)$, a convenient
representation for ${\bf f}$ is
\begin{eqnarray}
{\bf f} = U({\bf k},{\bf r}) 
\left( \begin{array}{cc} f_\uparrow({\bf k},{\bf r}) & 0 \\ 0 &
f_\downarrow({\bf k},{\bf r}) \end{array}\right) 
U^\dagger({\bf k},{\bf r})
\label{eq:matrix}
\end{eqnarray}
where
\begin{eqnarray}
U({\bf k},{\bf r}) =
\left( \begin{array}{cc}
 \cos(\theta/2) e^{-i\phi/2}  & -\sin(\theta/2) e^{-i\phi/2} \\
 \sin(\theta/2) e^{ i\phi/2}  &  \cos(\theta/2) e^{ i\phi/2}
\end{array} \right) 
\label{eq:unitary}
\end{eqnarray}
is the usual rotation matrix for spinors. We have suppressed the
${\bf k}$ and ${\bf r}$ dependence of $\theta$ and $\phi$ for
simplicity. Then, since the matrix ${\bf g}$ is related to ${\bf f}$ as
$g_{\sigma}$ is related to $f_{\sigma}$, we describe
\cite{BaymPethick:1991} the transport in
the spacer layer using the matrix analog of Eq.~(\ref{eq:be_fm}):
\begin{eqnarray}
{\bf v}_{\bf k} \cdot
{ \partial {\bf g}({\bf k},{\bf r}) 
\over \partial {\bf r} }
- e {\bf E} \cdot
{\bf v}_{\bf k} 
{\bf I} 
&=&
- {{\bf g}({\bf k},{\bf r}) - \overline{\bf g}({\bf r}) \over \tau}.
\label{eq:be_gen}
\end{eqnarray}
In this equation, {\bf I} is the $2 \times 2$ identity matrix and
$\tau$ is the relaxation time in the non-magnetic spacer.


\begin{thebibliography}{24}
\bibitem{waffle} 
This semiclassical approximation works for layers thinner than
relevant mean-free paths when both hot-electron effects due to
ballistic transport and interference effects due to multiple
reflections are not important.
\bibitem{sfCo} We do not include spin-flip scattering in the magnetic
layers because their thicknesses are much shorter than the spin
diffusion length. If the opposite were true, spin-flip scattering in
the Co (only) would be sufficient to polarize the current. In that
case, the transition region between the unpolarized leads and
saturation polarization in the Co would occur at the ends of the
magnetic layers.
\bibitem{precession}Spatial precession occurs when we superpose two
wave functions with the same energy but different wavevectors. The
more familiar temporal precession occurs when we superpose two wave
functions with the same wave vector but different energies.
\end{thebibliography}
\end{document}